\documentstyle[prl,aps,multicol,epsf,graphicx,xspace,cite,preprint]{revtex}

\newcommand{\critical}[1]{$#1_{\mathrm{c}}$\xspace}
\newcommand{\Tc}{\mbox{\critical{T}}\xspace}
\newcommand{\T}{T}

\newcommand{\etal}{{\sl et\,al.}\xspace}

\newcommand{\ybco}{YBa$_2$Cu$_3$O$_{7-\delta}$\xspace}
\newcommand{\lbco}{La$_{2-x}$Ba$_x$CuO$_4$\xspace}

\newcommand{\lcmo}{La$_{1-x}$Ca$_x$MnO$_3$\xspace}
\newcommand{\sto}{SrTiO$_3$\xspace}

\newcommand{\degree}{$^{\circ}$\xspace}

\newcommand{\dxxyy}{d$_{x^2-y^2}$\xspace}

\newcommand{\Vmeas}{\mbox{$V_{\mathrm{meas}}$}\xspace}
\newcommand{\Vgi}{\mbox{$V_{\mathrm{g}1}$}\xspace}
\newcommand{\Vgii}{\mbox{$V_{\mathrm{g}2}$}\xspace}

\newcommand{\Vgis}{\mbox{$V_{\mathrm{g}1}^{\star}$}\xspace}
\newcommand{\Vgiis}{\mbox{$V_{\mathrm{g}2}^{\star}$}\xspace}

\newcommand{\Vgb}{\mbox{$V_{\mathrm{gb}}$}\xspace}
\newcommand{\Vcr}{\mbox{$V_{\mathrm{cr}}$}\xspace}
\newcommand{\Icr}{\mbox{$I_{\mathrm{cr}}$}\xspace}

\newcommand{\Rgb}{\mbox{$R_{\mathrm{gb}}$}\xspace}

\begin{document}
\draft
\title{Electronic Transport through \ybco Grain Boundary Interfaces between 4.2\,K and 300\,K }
\date{\today}
\author{C.\,W. Schneider, S. Hembacher, G. Hammerl, R. Held, A. Schmehl, A. Weber,\\ T. Kopp, and J. Mannhart}
\address{
Experimentalphysik VI, Center for Electronic Correlations and
Magnetism, Institute of Physics, Augsburg University, D-86135
Augsburg, Germany\\ }

\date{\today}
\maketitle

\begin{abstract}
The current-induced dissipation in \ybco grain boundary tunnel junctions has been measured between
4.2\,K and 300\,K. It is found that the resistance of 45\degree (100)/(110) junctions decreases
linearly by a factor of four when their temperature is increased from 100\,K to 300\,K. At the
superconducting transition temperature \Tc the grain boundary resistance of the normal state and of
the superconducting state extrapolate to the same value.
\end{abstract}

%

Soon after the discovery of superconductivity in \lbco~\cite{Bednorz} it was realized that the
normal-state properties of the high-\Tc cuprates differ significantly from those of the low-\Tc
superconductors. The most prominent and still unexplained normal state properties of the high-\Tc
compounds are the linear temperature dependence of the in-plane resistivity in the optimally doped
compounds \cite{Tozer}, spin and charge inhomogeneities \cite{Orenstein}, and the pseudogap present
in the underdoped regime \cite{Timusk}.
\par

The complex electronic behavior of the high-\Tc materials is also reflected in the characteristics
of their interfaces.  For $T < \Tc$, for example, large-angle grain boundaries are excellent
tunnelling barriers \cite{Hilgenkamprmp}, a feature unknown from conventional superconductors.
Several mechanisms responsible for the electronic transport properties of grain boundaries and
other interfaces have been identified for the temperature range below \Tc, such as the \dxxyy
dominated order parameter symmetry, structural effects and space charge layers
\cite{Hilgenkamprmp}. In fact, grain boundaries operated below \Tc have been used to reveal
fundamental properties of the high-\Tc cuprates, such as the existence of Cooper pairs \cite{Gough}
or the unconventional order parameter symmetry \cite{Tsuei}.

As is the case for $T < \Tc$, we expect valuable information on the interfaces of the cuprates, or
even on the bulk materials themselves, to be contained in the tunneling characteristics of grain
boundaries for $T > \Tc$. However, only very few data are available for this temperature regime
\cite{Hilgenkamprmp, Ransley}. This lack of data arises from the difficulty to measure the grain
boundary current-voltage (\textit{I}(\Vgb)) characteristics free of voltages produced by the
bridges that are needed to contact the interfaces. By utilizing a difference-technique to subtract
these unwanted voltages, we have now succeeded in measuring the current-voltage characteristics of
\ybco grain boundaries for  $4.2\,\textrm{K} \le T \le  300\,\textrm{K}$. These studies provide
evidence that the normal state resistance of 45\degree (100)/(110) tilt grain boundaries is not
influenced by the onset of the superconducting transition. The experiments furthermore reveal that
for $T > \Tc$ the resistance of these grain boundaries decreases linearly with increasing
temperature.

The experiments were performed with bicrystalline \ybco films grown by pulsed laser deposition at
\mbox{760\degree{}C} in $0.25$\,mbar of O$_2$ to a typical thickness of 40--50\,nm. The
\sto-substrates contained symmetric and (100)/(110)-asymmetric 45\degree [001]-tilt grain
boundaries, specified to within 1\degree.  After deposition, the samples were cooled during two
hours in an O$_2$-pressure of $400$\,mbar. Because the deposition geometry was tuned to optimize
the homogeneity of the samples, \Tc varied by less than 0.5\,K across the wafers, while the
thickness variations were smaller than 10\,\%. For the four-point measurements, gold contacts were
structured photolithographically before the \ybco films were patterned by etching in a H$_3$PO$_4$
solution, or by dry etching with an Ar ion-beam. The \ybco bridges had widths and lengths between
the voltage probes of $\approx$ 2.5\,$\mu$m and 30\,$\mu$m, respectively (see Fig.~1), and their
\Tc varied less than 200\,mK. In addition, several samples were patterned into Wheatstone-bridge
configurations, of which each arm consists of a meander line containing 23 straight sections with
widths of $\approx$ 6\,${\mu}$m. Two of the meander lines were patterned such that each of the 23
sections crosses the boundary once.

The technique to measure precisely the grain boundary resistance \Rgb  below \Tc is based on
measuring the \textit{I}(\Vgb) characteristics of the Josephson junctions while applying a
microwave field large enough to suppress their critical current. For this purpose, we used low
frequency microwaves ($ <$ 10\,GHz) to avoid inaccuracies resulting from the  Shapiro steps.

All measurements were conducted in shielded rooms using motor-driven dip-sticks. The samples were
cooled at 4.2 K in liquid Helium, at 77 K in liquid nitrogen, and at other temperatures via He oder
N$_2$ vapor and thermal conduction across the sample holder. Immersion into the cooling liquids did
not affect the characteristics,  providing evidence that self-heating is insignificant under the
experimental conditions.

For $T > \Tc$ the transport properties of the interfaces cannot be measured with a straightforward
four-point technique because the voltage \Vmeas generated by a bridge straddling the boundary is
composed of two contributions: a) the grain boundary voltage \Vgb and b) the voltages \Vgi and
\Vgii caused by the resistive grains inside the bridge. This problem can be solved by subtracting
from \Vmeas the grain contributions \Vgi and \Vgii.  For this, \Vgi and \Vgii are obtained by
approximating them with the voltages \Vgis and \Vgiis generated by independent intragrain bridges
that are measured simultaneously. Not surprisingly, test measurements revealed variations of \Vgis
and \Vgiis of several percent as a function of the bridge location. Therefore, the intragrain
bridges were placed as closely as possible to the boundary bridge (see Fig.~1). The desired
measurement accuracy also determined the optimum size of the bridges. On the one hand the bridges
were patterned to be as small as possible so that they could be located close to the grain
boundary. On the other hand, their size was chosen to be large enough for the patterning-induced
scatter of the bridge aspect ratios to be insignificant. In samples optimized this way, the
voltages across the two bridges \Vgis and \Vgiis were identical to within 1\,\% at 100\,K and to
0.3\,\% for $T>200$\,K. Therefore the grain boundary voltage equals with the same accuracy $ \Vgb
\approx \Vmeas - (\Vgis+\Vgiis)/2$.

As introduced by Mathur~\etal \cite{Mathur} for studies of \lcmo bicrystals, and similar to the
work described in Ref.~\cite{Ransley}, we also patterned several samples into Wheatstone-bridges.
This approach has the advantage that the generated voltages are huge and that  \Vgis and \Vgiis are
subtracted from \Vmeas already during the measurement. Having performed measurements with such
bridges, we checked their balance by photolithographically cutting the Wheatstone-configurations,
to individually assess the resistances of the  meander lines. Because these studies revealed $T$
dependent balancing errors of 1--10\,\%, we preferred the three-bridge approach for measurements of
the $\Rgb(T)$) characteristics, while Wheatstone-bridges were chosen when a large signal was
required, as was the case for some studies of $I(V)$ characteristics.

To gain insight into the electronic transport mechanisms, we measured the $I(V)$ characteristics
between 4.2\,K and 300  K by using Wheatstone-bridges.  For $T>\Tc$ the characteristics are
non-linear on this large voltage scale, in particular below 150\,K (see Fig.~2). Analyzing the
non-linearity, we first note that the microstructure of the  grain boundaries is inhomogeneous down
to the unit cell level \cite{Hilgenkamprmp}. The $I(V)$ curves are therefore generated by large
numbers of microstructurally different channels connected in parallel. The behavior of the averaged
junctions is consistent with the one of a back-to-back Schottky contact as predicted by the band
bending model \cite{Mannhart98}. If tentatively described by a Simmons-fit \cite{Simmons}, the
averaged $I(V)$ characteristics correspond to heights and widths of hypothetical effective junction
barriers of $\approx$\,100 \,meV and 1--2 nm. Thus, these data suggest unusually small barrier
heights. They are, in particular, much smaller than the energy scale of grain boundary built-in
potentials measured by electron-holography of $\approx$\,2 eV \cite{Schofield}.

At small voltages, $\Vgb \lesssim $ 10\,mV, the non-linearity of the $I(V)$ curves is
insignificant, and  they are characterized by their ohmic resistance, which can be obtained by
selecting an appropriate voltage (\Vcr) or current (\Icr) criterion $\Vcr = \Icr \times \Rgb < $
10\,mV. For $T<\Tc$, due to the microwave irradiation, the $I(V)$ characteristics are linear, too,
even over a larger voltage range. The only exception is given by 45\degree boundaries at $T <
40$\,K, which will be discussed below. Applying the three-bridge technique, we were therefore able
to deduce the $\Rgb{(T)}$ dependence of  a given grain boundary by measuring in one temperature
sweep from 4.2\,K to 300\,K just one boundary bridge plus the two intragrain reference bridges.
Hereby one voltage or current criterion, typically $\Icr=100\,\mu$A, is used for all temperatures.

The $R(T)$-dependence of a 45\degree symmetric grain boundary measured this way is presented in
Fig.~3. The resistance-area product $\Rgb{A}$\,(77 K) $\approx$ $1\times10^{-8}$ $\Omega$cm$^2$
compares well with literature values \cite{Hilgenkamprmp}. Three remarkable features are displayed
by the temperature dependence of the resistance. First, the resistance reaches a maximum at
$\approx$ 30\,K (see also Fig.~4). This maximum is not displayed by symmetric 24\degree and
36\degree boundaries measured as reference samples. Second, around \Tc the resistance develops a
distinct peak structure. This peak, which is not associated with the microwave irradiation, was
found to vary from sample to sample, and to change with time over weeks. Except for this peak, the
grain boundary resistance of the normal state and of the superconducting state extrapolate within
the measurement accuracy of $\approx 2\,\%$ to the same value \Rgb(\Tc) = 10.7\,$\Omega$. This
temperature dependence is characteristic for all samples we have studied, except for few that
displayed at \Tc a resistance step of $\lesssim 5\%$. As this step was found to be not reproducible
and to increase for a given sample over weeks, we consider it to be an artifact resulting from
chemical reactions or diffusion. Third, between 100\,K and 300\,K the resistance decreases with
increasing temperature by a factor of four. The temperature dependence is hereby remarkably linear,
with a barely noticeable positive curvature (Fig.~3, upper inset). This is in striking contrast to
the linear resistance increase of the adjacent \ybco grains (Fig.~3, lower inset). In the
following, we will discuss the three features and their implications for the understanding of the
interfaces, beginning with the superconducting regime.

At low temperatures, the resistance of 45\degree grain boundaries changes non-monotonically with
$\T$, and even depends on the value of the tunneling current (see Fig.~4). This maximum and its
current bias dependence are consistent with the formation of a zero-bias anomaly by the faceted
45\degree junctions. At small bias currents \Vgb is of the order 1\,mV, and therefore sufficiently
small to probe the temperature dependent shape of the zero-bias conductance peak in the
superconducting energy gap. This zero-bias conductance peak is usually presumed to arise from
Andreev bound states generated by the \dxxyy order parameter symmetry of \ybco \cite{Hu}.

The  peak structure at 89 K provides evidence that at \Tc the resistance of the grain boundary
bridge is much larger than the resistance of the intragrain-bridges. The observed height of the
peak shown in Fig.~3 is consistent with a \Tc difference of the bridges of 150 mK. The shape of the
peak is influenced by the differences in the fluctuations of the intragrain bridges and the one
containing the tunnel junction. Unfortunately we cannot extract data on this effect from the peak,
as the spatial distribution of \Tc is unknown.

The grain boundary resistance values above and below \Tc extrapolate to the same value at the
transition temperature. The transition into the superconducting state is therefore found not to
affect the dissipation in the junctions at \Tc. This observation cannot be reconciled with the
predictions for surface charging effects in the hole superconductivity scenario introduced by
Hirsch and coworkers \cite{Hirsch1}. In this scenario, charges are separated in the superconducting
state in the vicinity of a grain boundary, but not above \Tc. This charge separation is expected to
change the electronic properties of the interfaces when passing through \Tc \cite {Hirsch2}. The
grain boundary data do not show this effect.

For $T> T_c$, the grain boundary resistance decreases linearly. This behavior cannot be explained
by elastic tunneling from the vicinity of a well-defined Fermi surface across standard Schottky
contacts with  $T$-independent barrier heights and widths ~\cite{Wolf}. There, a nonlinear and weak
$R_{\rm gb}(T)$ characteristic is expected for the parameter range derived above. A more detailed
elastic tunneling model has to include $T$ dependent Schottky barriers and to consider the spatial
inhomogeneity of the interface. To significantly alter the weak and nonlinear $T$ dependence both
effects demand the thermal energy $k_B\,T$ to exceed the energy scales of the barrier for large
parts of the interface areas. Hereby a very special energy profile of the barriers is required to
yield the linear $R_{\rm gb}(T)$ behavior over a large temperature range. This situation seems
unphysical to us. We note that inelastic tunneling via resonant impurity channels~\cite{Glazman}
cannot account for the linear temperature characteristic either. Considering the problems which
standard semiconductor models have in describing the observed interface characteristics, we now
turn our attention to the effect of the electronic correlations present in the cuprates.

As pointed out by Miller and Freericks~\cite{Freericks} electronic correlations strongly affect the
barrier properties of high-$T_c$ interfaces in the superconducting state. This is also true in the
normal conducting state. As suggested by the band bending model~\cite{Mannhart98}, the barrier is
an underdoped region, with the possible formation of antiferromagnetic fluctuations or even of
local magnetic moments. The resultant magnetic scattering is expected to increase the interface
resistance with decreasing temperature. Since the density of the magnetic moments is not known, the
$R_{\rm gb}(T)$ dependence cannot be determined yet. Besides correlations in the barrier,
electronic correlations in the bulk may also influence the transport through the interface. What
would be expected if, due to the grain boundary charging, the \ybco layers adjacent to the grain
boundary were underdoped, characterized by a pseudo gap and, possibly, by preformed pairs? The
spectral densities of the electronic excitations on both sides of the interface control the
tunneling process. It is only for the  45$^{\circ}$ grain boundaries under consideration, that the
nodal and antinodal directions of the pseudogap are directly coupled, and the tunneling current is
most strongly reduced. As the pseudo gap diminishes for increasing temperature, the boundary
conductivity is expected to increase. Here we note that the $R_{\rm gb}(T)$ curve is characterized
by one temperature scale, $T_0$, the extrapolated zero-resistance intercept. In the samples
investigated, $T_0\simeq 350$~K, which corresponds to the pseudo gap energy scale.

In summary, it is found that the resistance of 45\degree [001]-tilt \ybco grain boundary junctions
decreases linearly with increasing temperature for $\T>\Tc$. In the vicinity of \Tc the grain
boundary resistance and thus the dissipation are indistinguishable for $T<\Tc$ and $\T>\Tc$, in
contradiction to predictions of the theory of hole superconductivity. The $I(V)$ characteristics
between 4.2~K and 300~K are nonlinear, suggesting a tunneling barrier with an effective height and
width of $\approx$ 0.1 eV and $\approx$\, 1--2 nm, respectively. The linear $R_{\rm gb}(T)$
dependence is proposed to be caused by  correlation controlled tunneling.

The authors gratefully acknowledge helpful discussions with Y.~Barash, H.~Bielefeldt, M.~Blamire,
U.~Eckern, P.~J.~Hirschfeld, C.~Laschinger, J.~Ransley, A.~Rosch, D.~G. Schlom, and M.~Siegel. This
work was supported by the DFG through the SFB 484 and by the BMBF via project 13N6918A.
%


\begin{figure}[bpth]
\vspace{1cm}
%
\caption{\label{fig1}Optical micrograph of a three-bridge sample
used to measure \Rgb. The 45\degree grain boundary (gb) is
indicated by the arrows.}
\end{figure}

\begin{figure}[bpth]
%
\caption{\label{fig2}Current-voltage characteristics of a
Wheatstone-bridge containing 23 junctions with a 45\degree
asymmetric boundary in a 50\,nm thick \ybco film. The voltage $V$
corresponds to the voltage across all junctions, $V=23\times
\Vgb$, the current $I$ is twice the current flowing through one
meander line.}
\end{figure}

\begin{figure}[bpth]
%
\caption{\label{fig3} Measured temperature dependence of the
resistance of a 45\degree symmetric grain boundary in a 40\,nm
thick \ybco film ($\Icr=100\,\mu$A). The peak reaches a maximum of
25\,$\Omega$ at 88.3\,K. The insets show d$R$/d$T(T)$ for the
grain boundary and the corresponding $R(T)$ curve of one grain
located next to the boundary. }
\end{figure}

\begin{figure}[bpth]
%
\caption{\label{fig4}Temperature dependence of the resistance of a
45\degree symmetric grain boundary in a 40\,nm thick \ybco film
measured in the three-bridge configuration. The inset shows the
temperature dependent resistance of a 45\degree asymmetric grain
boundary in a 50\,nm thick film measured in a Wheatstone-bridge
configuration.}
\end{figure}

\newpage

~\vspace{5cm}
\center\includegraphics[width=0.75\columnwidth,clip=true]{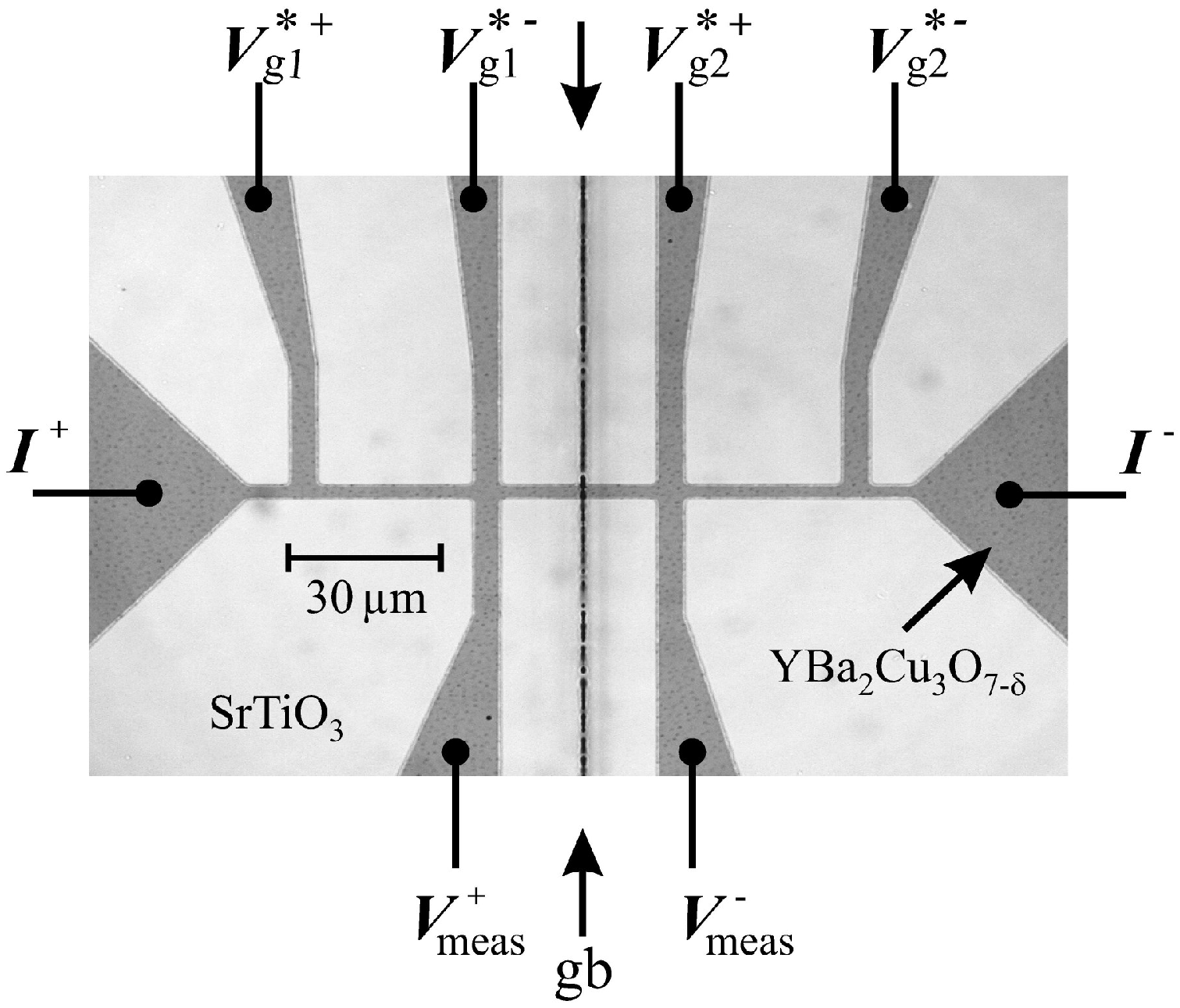}
\vspace{0.5cm} \vfill \raggedleft{\sl Schneider \etal, Figure 1}

\newpage

~\vspace{5cm}
\center\includegraphics[width=0.75\columnwidth,clip=true]{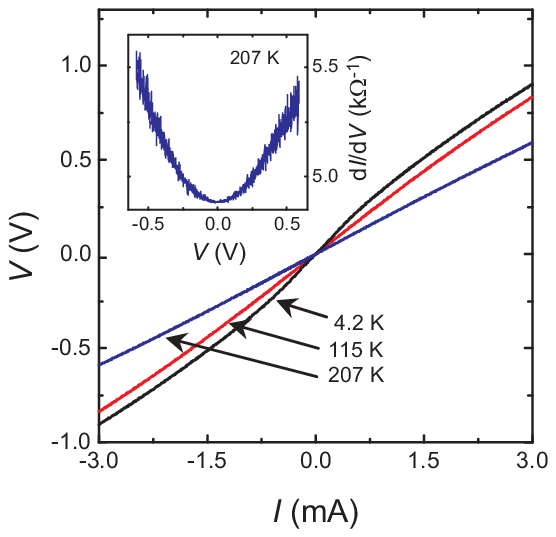}
\vspace{0.5cm} \vfill \raggedleft{\sl Schneider \etal, Figure 2}

\newpage

~\vspace{5cm}
\center\includegraphics[width=0.87\columnwidth,clip=true]{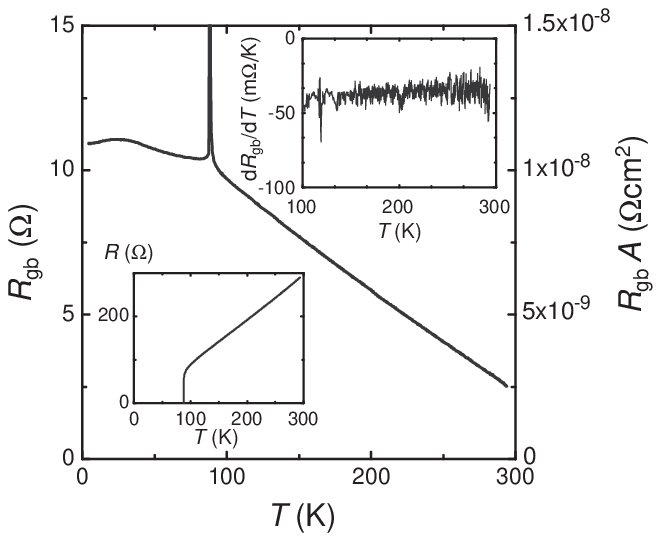}
\vspace{0.5cm} \vfill \raggedleft{\sl Schneider \etal, Figure 3}

\newpage

~\vspace{5cm}
\center\includegraphics[width=0.93\columnwidth,clip=true]{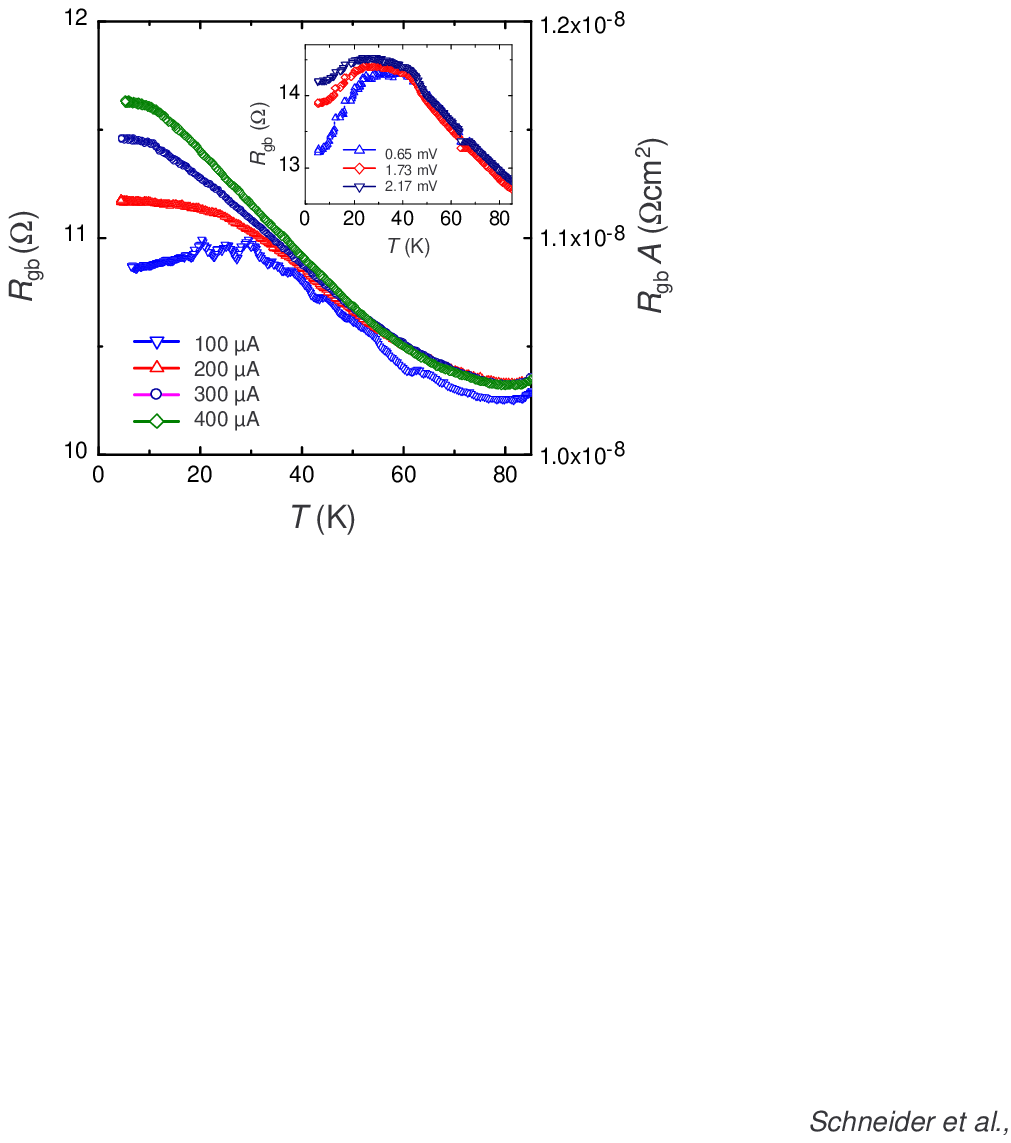}
\vspace{0.5cm} \vfill \raggedleft{\sl Schneider \etal, Figure 4}


\begin{references}
%
\bibitem{Bednorz} J.\,G. Bednorz and K.\,A. M\"{u}ller, Z. Phys. B {\bf 64}, 189 (1986).
%
\bibitem{Tozer} S.W.~Tozer \etal, \prl {\bf 59}, 1768 (1987).
%
\bibitem{Orenstein} J. Orenstein and A.\,J. Millis, Science {\bf 288}, 468 (2000).
%
\bibitem{Timusk} T. Timusk and B. Statt, Rep. Prog. Phys {\bf 62}, 61 (1999).
%
\bibitem{Hilgenkamprmp} H. Hilgenkamp and J. Mannhart, \rmp{\bf 74}, 485 (2002).
%
\bibitem{Gough} C.\,E.~Gough \etal, Nature {\bf 326}, 855 (1987).
%
\bibitem{Tsuei} C.\,C. Tsuei and J.\,R. Kirtley, \rmp{\bf 72}, 969 (2000).
\bibitem{Ransley} J.\,H.\,T.~Ransley \etal, IEEE Trans. Appl. Supercond. {\bf 13}, 2886 (2003).
%
\bibitem{Mathur} N.\,D. Mathur \etal,
Nature {\bf 387}, 266 (1997).
%
\bibitem{Mannhart98} J.~Mannhart and H.~Hilgenkamp, Mat. Sci. Eng. {\bf B56}, 77 (1998).
%
\bibitem{Simmons} J.\,G.~Simmons, J. Appl. Phys. {\bf 34}, 1828 (1963).
%
\bibitem{Schofield} M.\,A.~Schofield \etal, \prb{\bf 67}, 224512 (2003).
%
\bibitem{Hu} C.-R. Hu, \prl {\bf72}, 1526 (1994).
%
\bibitem{Hirsch1} J.\,E.~Hirsch and F. Marsiglio, \prb {\bf 39}, 11515 (1989).
%
\bibitem{Hirsch2} J.~Hirsch, Phys. Lett. A {\bf 281}, 44 (2001).
%
\bibitem{Wolf} E.~L.~Wolf, {\it Principles of Electron Tunneling
Spectroscopy} (Clarendon Press, Oxford, 1985).
%
\bibitem{Glazman} L.\,I.~Glazman and K.\,A.~Matveev, Sov. Phys. JETP {\bf 67}, 1276 (1988); err: {\bf 49}, 659 (1989).
%
\bibitem{Freericks} P.\,~Miller and J.\,K.~Freericks, Journ. of Phys. {\bf 13}, 3187 (2001).




\end{references}
\end{document}